\documentclass[aps,prl,reprint,superscriptaddress]{revtex4-2}
\usepackage{float}
\usepackage{siunitx}
\usepackage{graphicx}
\usepackage{dcolumn}
\usepackage{bm}

\usepackage[utf8]{inputenc} 
\usepackage[T1]{fontenc}   
\usepackage{amsmath}       
\usepackage{amssymb}       
\usepackage{graphicx}      
\usepackage{hyperref}      
\usepackage{xcolor}        
\usepackage{braket}
\usepackage{subcaption}

\begin{document}

\preprint{APS/123-QED}

\title{Observation of intrastate and interstate facilitation between Rydberg S, P and D levels}

\author{Bleuenn Bégoc}
\affiliation{CNR-INO, via G. Moruzzi 1, 56124 Pisa, Italy}
\affiliation{Dipartimento di Fisica dell’Università di Pisa, Largo Pontecorvo 3, 56127 Pisa, Italy}

\author{Sukhjit P. Singh}
\affiliation{Dipartimento di Fisica dell’Università di Pisa, Largo Pontecorvo 3, 56127 Pisa, Italy}

\author{Giovanni Cichelli}
\affiliation{Dipartimento di Fisica dell’Università di Pisa, Largo Pontecorvo 3, 56127 Pisa, Italy}

\author{Roberto Franco}
\affiliation{Dipartimento di Fisica dell’Università di Pisa, Largo Pontecorvo 3, 56127 Pisa, Italy}

\author{Oliver Morsch}
\affiliation{CNR-INO, via G. Moruzzi 1, 56124 Pisa, Italy}
\affiliation{Dipartimento di Fisica dell’Università di Pisa, Largo Pontecorvo 3, 56127 Pisa, Italy}

\date{\today}

\begin{abstract}

We report experimental results on Rydberg facilitation, whereby Rydberg levels can be excited off-resonantly in the presence of a nearby Rydberg atom because of Rydberg-Rydberg interactions, for high-lying $S$, $P$ and $D$ levels in rubidium. Facilitation is detected both through an enhancement of the number of excited atoms for off-resonant excitation (either blue or red detuning) and a positive Mandel $Q$ parameter indicating correlated excitation events (super-Poissonian counting statistics). We also calculate the pair-state potentials for the Rydberg states involved and find that our experimental results agree with the expected facilitation conditions for repulsive potentials (blue detuning) and attractive potentials (red detuning), with $P$ and $D$ states exhibiting facilitation on both sides of the resonance. Finally, we investigate inter-state facilitation between two different Rydberg levels (70 $S$ and 70 $P$).

\end{abstract}

\maketitle

\maketitle


\section{Introduction}

Rydberg excitations in ultracold atoms have become a rich and versatile model system for strongly correlated phenomena in recent years, in various applications such as quantum computing and quantum simulations \cite{spininteractions, Saffman2010,armin,Griffiths:1492149}. For example, recent studies have exploited Rydberg atoms to simulate epidemic growth \cite{Tobpap} and in realising quantum cellular automata (QCA) \cite{Tobpap2}.
Their strong long-range interactions constitute a key resource for quantum computing applications. In particular, in analog quantum simulators, these interactions enable the implementation of spin Hamiltonians such as the Ising and XY models \cite{Saffman2010, browaeys2020, weimer2010, bernien2017}.

The strong van der Waals and dipole-dipole interactions between high-lying Rydberg states lead to rich many-body dynamics and collective effects. One of the most prominent phenomena is the Rydberg blockade, where the excitation of a single atom to a Rydberg state inhibits the excitation of nearby atoms within a characteristic blockade radius, due to the energy shifts induced by the interaction \cite{Bloch,Low2012,blockade,superblock,blocknature}. Conversely, under off-resonant driving, these interactions can also lead to facilitated excitation, where the presence of a Rydberg excitation shifts nearby atoms into resonance, promoting the creation of correlated excitation clusters \cite{kineticconstraints,facvapor,Oliver,fac,fac2dlattice,periodicfac,bluefac,rydberglifetime,introentangle}.
So far, Rydberg facilitation has been studied mostly between atoms in the same $nS$ Rydberg state with repulsive interactions between them \cite{fac,Tob}. We will call facilitation between the same Rydberg state intrastate facilitation.
However, to our knowledge, facilitation in other interaction regimes, such as attractive or mixed interactions, remains largely unexplored. Furthermore, interstate facilitation, where excitation in one Rydberg level facilitates excitation of a different Rydberg level of another atom, has not been experimentally demonstrated to our knowledge. In this work, we experimentally investigate facilitated excitation across different intrastate interaction regimes, including repulsive, attractive, and mixed cases. In addition, we demonstrate interstate facilitation between atoms in two different Rydberg levels.
\\

\section{Facilitated excitation}

\begin{figure*}[!t]
\includegraphics[width=1\textwidth]{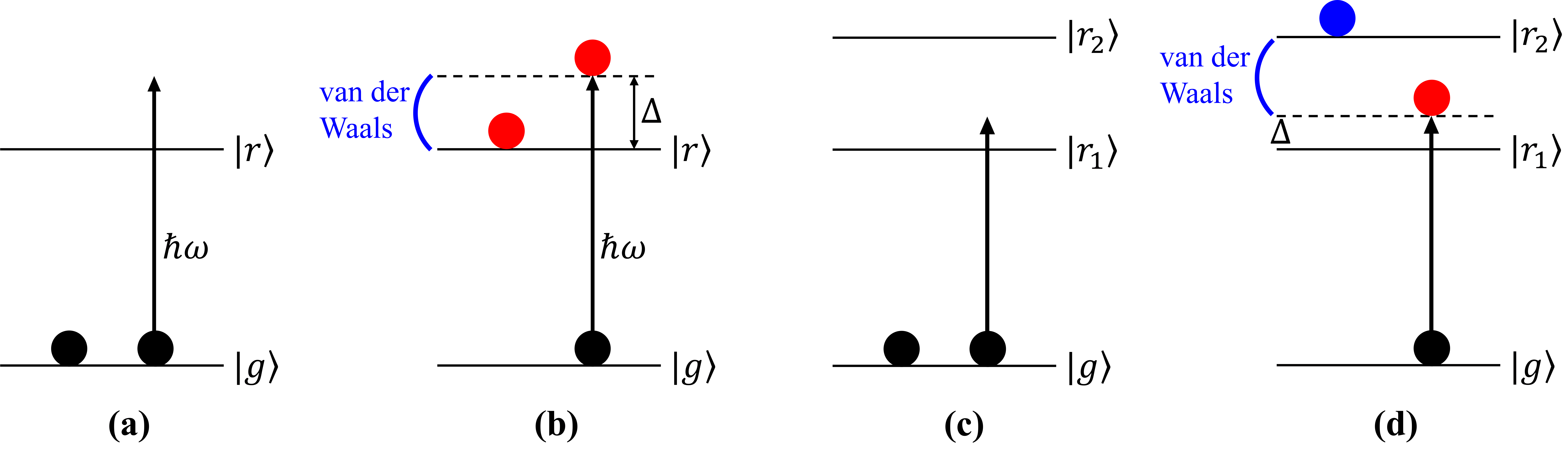}
\captionsetup{
  justification=raggedright,
  singlelinecheck=false
}
\caption{Intrastate and Interstate facilitation process between two atoms.  Intrastate facilitation (a), Without a nearby seed atom in a Rydberg state, the off-resonant laser radiation does not excite the atom to the Rydberg state. (b) When an already excited atom is present, it shifts the energy level of a nearby atom via van der Waals interaction, allowing off-resonant excitation. Interstate facilitation (c), Without a nearby atom in the Rydberg state in $\ket{r_2}$, no excitation occurs. (d) With a nearby atom in the Rydberg state $\ket{r_2}$, facilitated excitation of the other atom to $\ket{r_1}$ is possible.} 
\label{fig:fac}
\end{figure*}

Facilitated excitation arises from the strong van der Waals interactions between atoms excited to high-lying Rydberg states. 
When an initial "seed" atom is excited to a Rydberg state, the Rydberg levels of the surrounding atoms in the sample experience a shift due to the van der Waals interaction. If the excitation laser is detuned from the single-atom resonance, this interaction can shift the levels of surrounding atoms closer to resonance, as illustrated in the Fig.\ref{fig:fac}(a)-(b).
At a specific distance called the facilitation radius $r_{\mathrm{fac}}$, the interaction shift matches the laser detuning $\Delta$, and facilitated excitation occurs. The facilitation condition is given by:
\begin{equation}
\Delta = \frac{V(r_{\mathrm{fac}})}{\hbar}
\label{eq:fac_distance}
\end{equation}

For intrastate facilitated excitation, the interaction is of van der Waals type, with $V(r_{\mathrm{fac}})=\frac{C_6}{r_{\mathrm{fac}}^6}$ where $C_6$ is the van der Waals interaction coefficient.

For interstate processes, provided that the system is far from any Förster resonance, dipole-dipole interactions can be treated perturbatively and give rise to an effective van der Waals interaction with the same $1/r^6$ scaling.

The presence of one Rydberg excitation thus defines a spherical shell of radius $r_{\mathrm{fac}}=(\frac{C_6}{\hbar\times\Delta})^{1/6}
$ and thickness $\delta r_{\mathrm{fac}} = r_{\mathrm{fac}} \frac{\gamma}{6\Delta}$ in which atoms can be resonantly excited to the Rydberg state, 

Here, $\gamma$ is the effective linewidth of the Rydberg transition, which can be limited by the lifetime of the Rydberg state or by other processes like laser jitter. In our experiments, $\gamma$ is around $2\pi \times 400\,\mathrm{kHz}$ and $\Delta$ is the detuning of the laser from the atomic resonance.
If the system evolves for a sufficiently long duration, this mechanism can initiate an avalanche of excitations, generating clusters of Rydberg atoms.\\
\\
For a duration that allows the facilitation avalanche to occur, one primary signature of facilitated excitation is the asymmetric dependence of the total number of Rydberg excitations on the laser detuning.  
For repulsive interactions, there will be an asymmetry towards positive detuning, whereas attractive interactions will lead to an asymmetry towards negative detuning. 
This asymmetry becomes more pronounced as the duration of the excitation pulse increases, allowing the facilitation avalanche to grow.
The resonant excitation rate in the facilitation regime can be approximated by \cite{Oliver}:
\begin{equation}
\Gamma_{\mathrm{fac}} = \frac{\Omega^2}{2\gamma},
\label{equgammafac}
\end{equation}
where $\Omega$ is the effective Rabi frequency and $\gamma$ is the decoherence rate, which in our experiments is dominated by the laser jitter. In this paper we limit ourselves to parameters for which single excitation and facilitation events are incoherent and hence can be described by a rate equation.

Generally, the excitation rate for a non-interacting atoms follows the Lorentzian function \cite{fox2006quantum}:
\begin{equation}
\Gamma_{\mathrm{exc}} = \frac{\Omega^2}{\gamma} \times \frac{1}{\frac{\Delta^2}{\gamma^2} + 1},
\label{equgamma}
\end{equation} 
In the absence of interactions, excitation is strongly suppressed for $\Delta\gg \gamma$.
\\
Another signature of facilitated excitation is obtained from the counting statistics of Rydberg excitations via the Mandel $Q$ parameter \cite{gunter2013facilitation}:
\begin{equation}
Q=\frac{\langle (\Delta N)^2 \rangle}{\langle N \rangle}-1
\end{equation} 
where \(\langle N \rangle\) denotes the mean number of excitations and \(\langle (\Delta N)^2 \rangle\) their variance.
For a Poissonian process, characteristic of independent excitation events, the variance equals the mean,
$\langle (\Delta N)^2 \rangle = \langle N \rangle$, yielding $Q = 0$.

This regime is expected for resonant excitation at short excitation times, for which few Rydberg excitations are created at large mean distances, and hence interactions play a negligible role.

For longer excitation times, two distinct phenomena can occur. For zero detuning (resonant excitation), excitation of nearby atoms is suppressed. This Rydberg blockade \cite{kineticconstraints,facvapor,Oliver,fac,fac2dlattice,periodicfac,bluefac,rydberglifetime,introentangle} leads to reduced number fluctuations, $\langle (\Delta N)^2 \rangle < \langle N \rangle$, and consequently to negative values of the Mandel parameter ($Q < 0$), reflecting sub-Poissonian statistics and anti-correlated excitation events. In contrast, facilitated excitation leads to the formation of correlated excitation clusters. Once a Rydberg atom is present, additional excitations are enhanced at specific distances, resulting in large shot-to-shot fluctuations in the total number of excitations.
In this regime, the variance exceeds the mean, $\langle (\Delta N)^2 \rangle \gg \langle N \rangle$, yielding large positive values of the Mandel parameter
($Q \gg 1$). Such super-Poissonian statistics constitute a clear signature of facilitation-driven avalanche dynamics.

In this paper, we investigate Rydberg facilitation between atoms in the same Rydberg state, referred to as intrastate facilitation, as illustrated in Fig.~\ref{fig:fac}(a)-(b), as well as interstate facilitation, where excitation is facilitated between two different Rydberg levels, as shown in Fig.~\ref{fig:fac}(c)-(d).
In this scenario, a pre-existing Rydberg seed atom in state \( \ket{r_2} \) induces an energy shift of the state \( \ket{r_1} \) via the van der Waals interaction. If the laser is detuned by an amount matching this interaction shift, a nearby atom at the facilitation distance can be resonantly excited to \( \ket{r_1} \). This excitation would not occur without the presence of the initial seed atom, as evident when comparing with Fig.~\ref{fig:fac}(c).
\\


\begin{figure*}[!t]
\includegraphics[width=0.75\textwidth]{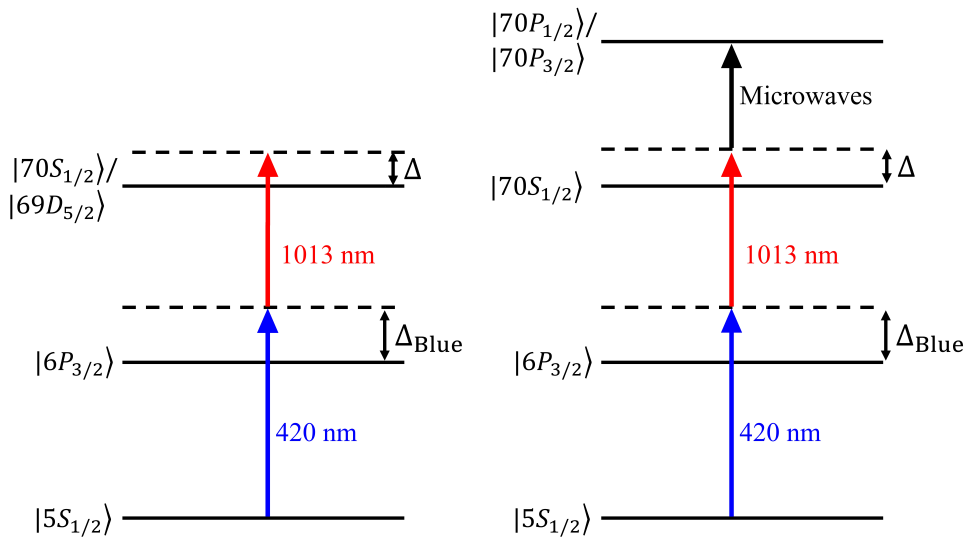}
\captionsetup{
  justification=raggedright,
  singlelinecheck=false
}
\caption{Excitation energy level sketch. Left: Two level system: Excitation into $\ket{70S_{1/2}}$ or $\ket{69D_{5/2}}$ states. Right: Three level system: Excitation into $\ket{70P_{1/2}}$, $\ket{70P_{3/2}}$ via $\ket{70S_{1/2}}$}
\label{fig:fac2}
\end{figure*}

\section{Method to observe facilitated excitation}

The experiments were performed with clouds of \(^{87}\)Rb atoms in a magneto-optical trap (MOT). The peak atomic density in the MOT was \(1 \times 10^{10} \,{\rm cm}^{-3}\), corresponding to an average interatomic distance of 4.6~\si{\micro\meter} in the center of the MOT.
Fig. \ref{fig:fac2} illustrates Rydberg excitation to states \( \ket{70S_{1/2}} \) and \( \ket{69D_{5/2}} \) via a two-photon transition using co-propagating laser beams at 420 nm and 1012 nm, via the \( \ket{6P_{3/2}} \) state with a detuning of between 350 MHz (for $\ket{70S_{1/2}}$) and 1 GHz (for $\ket{69D_{5/2}}$) to avoid populating that state.

Additionally, three-photon excitation was used to access \( \ket{70P_{1/2}} \) and \( \ket{70P_{3/2}} \) states via microwave coupling at 10.369 GHz for \( \ket{70P_{1/2}} \) and 10.654 GHz for \( \ket{70P_{3/2}} \) with an additional intermediate state ($70S_{1/2}$) \cite{townes1955microwave}.
For resonant excitation ($\Delta = 0$), using Eq.~(\ref{equgammafac}) with $\gamma = 251 \times 10^4~\mathrm{s^{-1}}$ for the $70S_{1/2}$ state due to the laser jitter and a two-photon Rabi frequency $\Omega_{\mathrm{2ph}} = 2\pi\times108 \times 10^3~\mathrm{kHz}$, the single-atom excitation rate is $\Gamma \simeq 78.5\times 10^3~\mathrm{s^{-1}}$, corresponding to a single-atom excitation time of 12.8 $\si{\micro\second}$.

To ensure excitation beyond the single-atom regime, we apply excitation pulses of up to $100~\si{\micro\second}$, allowing sufficient time for multiple facilitated Rydberg excitations and enabling the facilitated excitation process to propagate through the atomic cloud.

Detection of Rydberg excitations is performed via field ionization followed by time-resolved ion counting using a channeltron detector. We then calculate the mean number of detected Rydberg atoms and their shot-to-shot fluctuations to extract the Mandel \(Q\) parameter, averaging over 100 sequences for each measurement point.

\section{Results}

\subsection{Intrastate facilitation}

We first investigate intrastate facilitation between atoms in the same Rydberg state, using four different Rydberg levels: $\ket{70S_{1/2}}$, $\ket{69D_{5/2}}$, $\ket{70P_{1/2}}$  and $\ket{70P_{3/2}}$.\\
\\
To quantify the conditions for facilitation, we compute the pair-state interaction energies as a function of interatomic distance using the python library ARC \cite{ARC2017}.

In principle, the interaction between two atoms depends on the different combinations of the magnetic quantum numbers $(m_{J1}, m_{J2})$. 

In a MOT, the quadrupole magnetic field does not define a unique quantization axis. As a result, with linearly

\begin{figure*}[!t]
  \centering
    \includegraphics[width=1.01\linewidth]{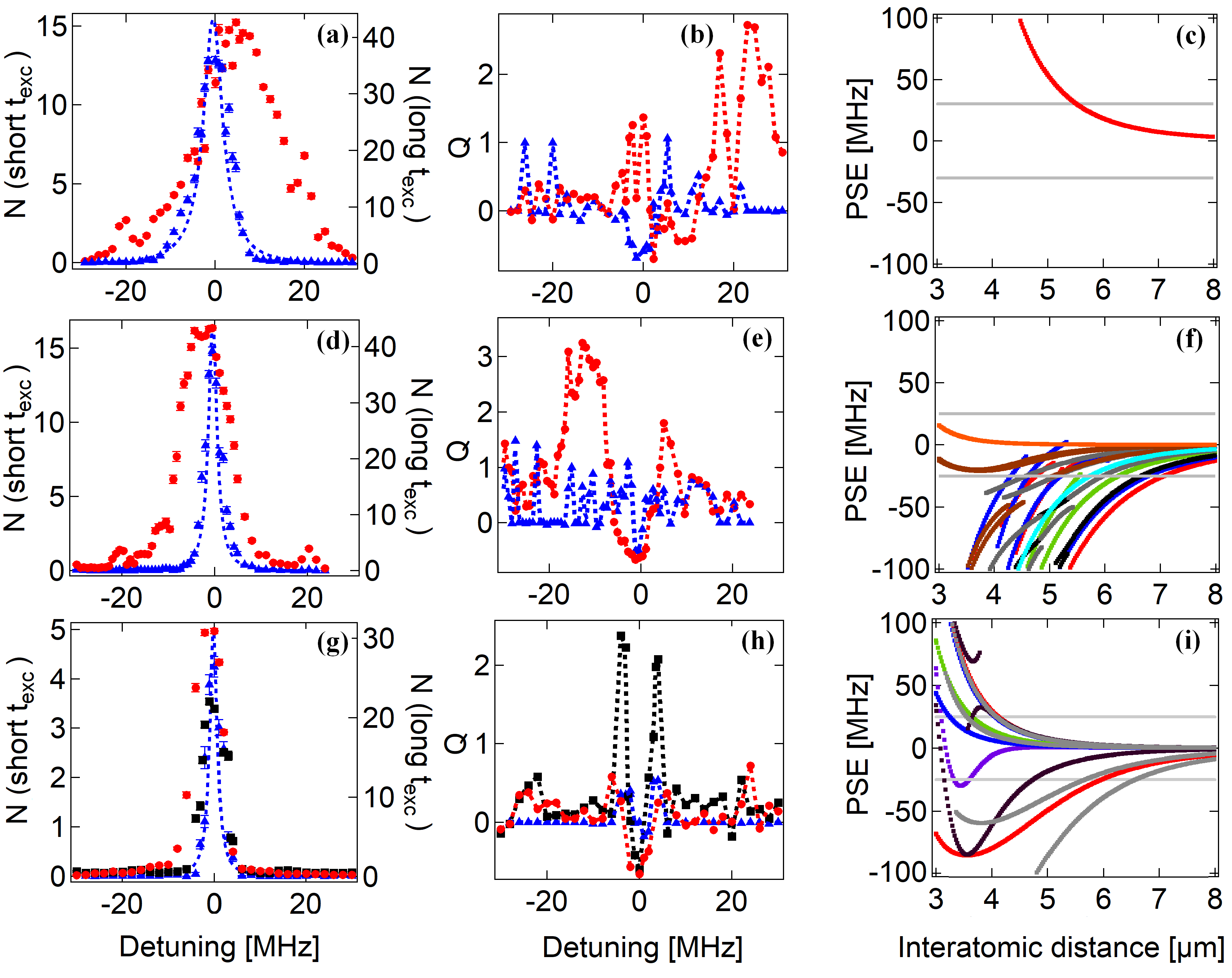}
\captionsetup{
  justification=raggedright,
  singlelinecheck=false
}
\caption{
Intrastate facilitation for different Rydberg states.
$70S_{1/2}$ (top row (a-c)); $69D_{5/2}$ (middle row (d-f)); $70P_{1/2}$ and $70P_{3/2}$ (bottom row (g-i)).
Left column (a)(d)(g): Number of excited Rydberg atoms ($N$) as a function of detuning, for various excitation durations $t_{\text{exc}}$. 
Blue triangles: short $t_{\text{exc}}$ is 0.3 $\si{\micro\second}$ for $70S_{1/2}$, 1 $\si{\micro\second}$ for $69D_{5/2}$ and 4 $\si{\micro\second}$ for $70P_{3/2}$ (single-atom regime) with Lorentzian fit (dashed blue line). Data in the single-atom regime for the $70P_{1/2}$ state are similar to those for the $70P_{3/2}$ state and not shown here for clarity. red circles: long $t_{\text{exc}}$ 100 $\si{\micro\second}$ (facilitation regime). Bottom row for $70P_{1/2}$; black rectangles: 100~$\si{\micro\second}$ for $70P_{3/2}$.
Middle column (b)(e)(h): Mandel $Q$ parameter as a function of detuning. The dashed lines connect the data points to guide the eye.
Right column (c)(f)(i): Interaction potentials between Rydberg atoms (Pair-State Energy (PSE)) as a function of interatomic separation.
The grey lines mark the range of detunings explored experimentally.
Red line: both atoms in the same sublevel with $m_j = 1/2$ (i) for $70P_{3/2}$. (a) for $70S_{1/2}$ also $m_j = 1/2$ and $m_j =-1/2$
Blue line: same sublevels $m_j = 3/2$ (i) for $70P_{3/2}$.
Green line: asymmetric pair states with $m_j = 1/2$ and $m_j = 3/2$ (i) for $70P_{3/2}$.
Purple line: $70P_{1/2}$ state with both atoms in $m_j = 1/2$.
Black line:  $m_j = 1/2$ and $m_j =-1/2$ (i) for $70P_{1/2}$.
Grey line:  (f) $m_j = 3/2$ and $m_j =-3/2$ and (i) $m_j = -1/2$ and $m_j =+3/2$ for $70P_{3/2}$.
Brown line: $m_j = \pm 5/2$ and $m_j =\mp 3/2$.
Cyan line:  $m_j = 5/2$ and $m_j =5/2$.
Orange line:  $m_j = -5/2$ and $m_j =5/2$. (f) The sublevels $m_j = \pm 5/2$ and $m_j = \mp 1/2$, $m_j = \pm 3/2$ and $m_j = \mp 1/2$ are not shown, as their behavior is similar to that of the sublevels $m_j = 3/2$ and  $m_j = 1/2$ and $m_j =-1/2$, in order to improve the readability of the graph.
}
\label{fig:1}
\end{figure*}

polarized lasers (420 and 1013 nm), several Zeeman sublevels of the Rydberg state all sublevels are excited  simultaneously.
The interaction therefore corresponds to an average over all $m_J$ states.
Due to the symmetry between $m_J$ and $-m_J$, this effective interaction can be well approximated by considering a single representative case, typically $(m_{J1},m_{J2})$ and not $(-m_{J1},-m_{J2})$ when calculating the $C_6$ coefficient.

The interaction between two atoms in the $\ket{70S_{1/2}}$ state is purely repulsive and follows a van der Waals potential of the form $V(r) = C_6/r^6$, with $C_6 = h \times 869.7~\text{GHz}\,\si{\micro\meter}^6$ [Fig.~\ref{fig:1}(c)]. 
The facilitation condition is satisfied when the laser detuning $\Delta$ matches the interaction energy shift at a given separation, for example at a facilitation radius of 
$r_{\text{fac}} \approx \num{5.5}\,\si{\micro\meter}$ for a positive detuning of $\Delta = +25$ MHz.

For the $\ket{69D_{5/2}}$ state, the interaction potential is attractive for most magnetic sublevels, leading to negative interaction energies at short distances [Fig.~\ref{fig:1}(f)]. Multiple pair potentials are observed due to different combinations of Zeeman sublevels, resulting in a facilitation shell that supports excitation at negative detuning.

\begin{figure*}[!t]
\centering
    \includegraphics[width=1.01\linewidth]{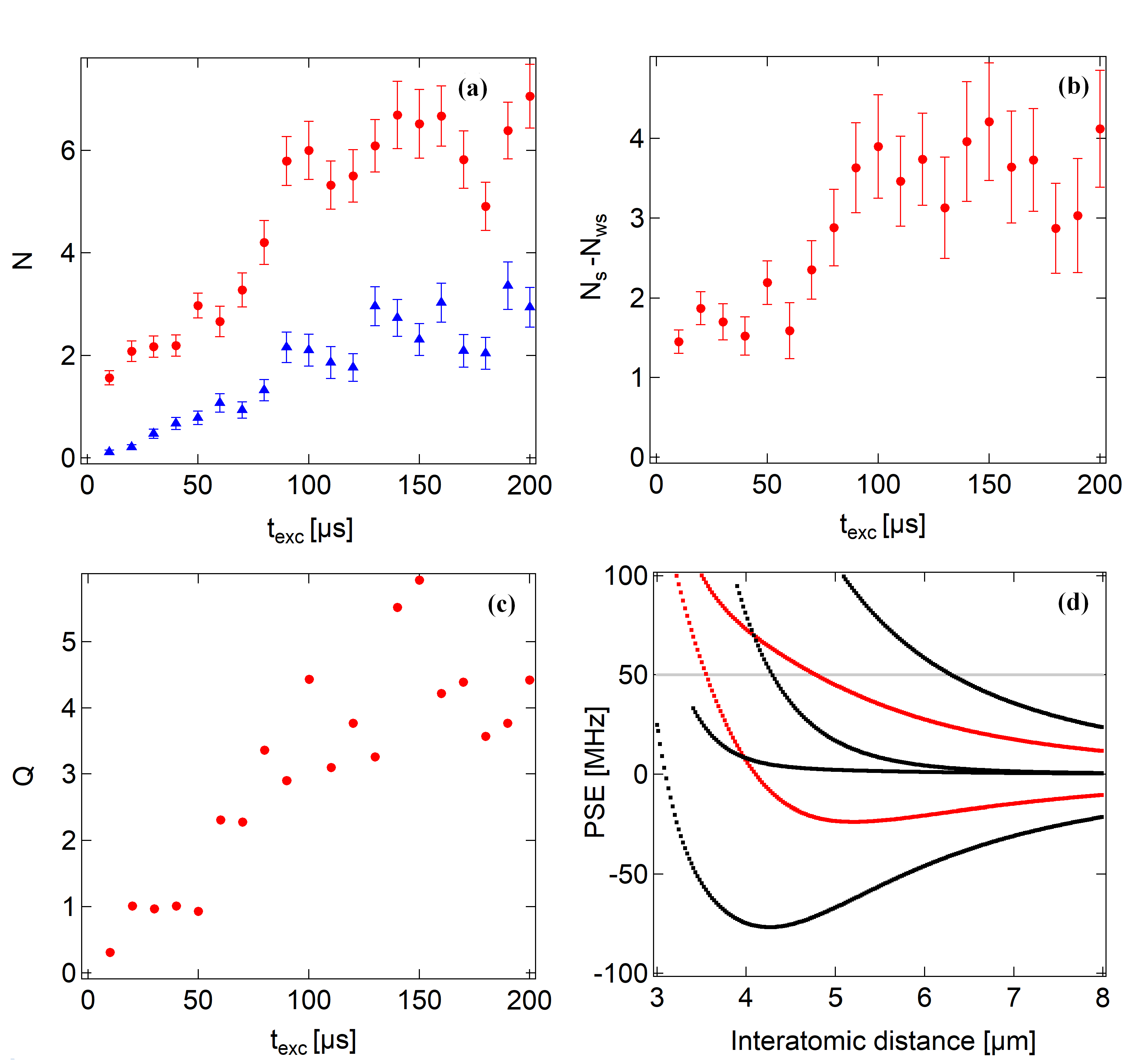}
\captionsetup{
  justification=raggedright,
  singlelinecheck=false
}
\caption{Interstate facilitation for $70S_{1/2}$-$70P_{1/2}$.
(a) Number of excited Rydberg atoms ($N$) as a function of excitation time $t_{\text{exc}}$. Red circles: initial seed in higher Rydberg level $70P_{1/2}$; blue triangles: no seed; green squares: Mandel parameter with an initial seed. (b) Difference of N with seed ($N_{S}$) and without seed ($N_{WS}$). (c)  Mandel parameter with an initial seed a function of excitation time $t_{\text{exc}}$ (d) Interaction potentials between atoms in the two Rydberg states (Pair-State Energy PSE) as a function of interatomic separation. Red line :  both atoms in the same sublevel with mj = 1/2. Black line : one atom in the sublevel 1/2, the other one in -1/2. The grey line indicates the experimental detuning.}
\label{fig:2}
\end{figure*}

For off-resonant excitation, by contrast, the $\ket{70P_{1/2}}$ and $\ket{70P_{3/2}}$ states exhibit a more complex behavior. Depending on the combination of magnetic quantum numbers $m_j$ and interatomic separation, the interaction can be either attractive or repulsive [Fig.~\ref{fig:1}(i)]. As a result, facilitated excitation is expected to occur at both positive and negative detunings.\\
\\

Figures~\ref{fig:1}(a),~\ref{fig:1}(d) and~\ref{fig:1}(g) show the mean number of Rydberg excitations $N$ as a function of laser detuning for two different excitation durations: a short pulse (between 0.3 and $4~\si{\micro\second}$, blue triangles) and a longer pulse ($100 \,\si{\micro\second}$, red symbols). 

For short pulses, the excitation profile follows a symmetric Lorentzian shape, consistent with the single-atom excitation regime. This symmetry arises because the excitation process is dominated by the excitation of individual atoms, with minimal contributions from interactions or multi-atom effects.
For longer excitation times (red circles: $t_{\text{exc}} = 100~\si{\micro\second}$), off-resonant excitations become visible, resulting in an asymmetric line shape, consistent with facilitated excitation. For the $\ket{70S_{1/2}}$ state, the excitation rate is strongly enhanced at positive detuning, in agreement with the repulsive interaction potential (we note here that the small peaks visible at $\pm$ 20 MHz detuning in Fig.\ref{fig:1} (a,d,g)  are artefacts due to sidebands on the 420 nm laser). Conversely, the $\ket{69D_{5/2}}$ state shows a similar enhancement at negative detuning, where facilitation is energetically favored due to attractive interactions. Moreover, some broadening is also observed for positive detuning, which is likely due to the repulsive interaction channel associated with the sublevels $-\frac{5}{2}, \frac{5}{2}$ and $\pm\frac{5}{2}, \mp\frac{3}{2}$.

The $\ket{70P_{1/2}}$ and $\ket{70P_{3/2}}$ states exhibit excitation enhancement on both sides of resonance, consistent with the mixed nature of the pair-state potentials.\\
\\

To further confirm the presence of interaction-induced facilitation dynamics, we analyse the counting statistics of the number of Rydberg excitations via the Mandel $Q$ parameter [Fig.~\ref{fig:1}(b), \ref{fig:1}(e), \ref{fig:1}(h)]. For resonant excitation in the single-atom regime, the $Q$ parameter remains close to zero, indicating a nearly Poissonian distribution. However, in the facilitation-dominated regime, the $Q$ parameter increases significantly at detunings corresponding to facilitated excitation, reflecting strong fluctuations and correlations among excitation events.

In the facilitation-dominated regime, for the $\ket{70S_{1/2}}$ state, we observe $Q > 2$ at positive detuning, while the distribution is Poissonian for negative detuning. The $\ket{69D_{5/2}}$ state exhibits the inverse behaviour, with values of the $Q$ parameter up to 3 at negative detuning as well as a small peak at positive detuning of around 5 MHz, consistent with the slight broadening of the line for positive detuning and the presence of repulsive pair-state potentials. The $\ket{70P_{1/2}}$ and $\ket{70P_{3/2}}$ states show positive $Q$ values on both sides of the resonance, confirming the presence of facilitation for positive and negative detuning.

\subsection{Interstate facilitation}

We also investigated interstate facilitation between the Rydberg states $\ket{70P_{1/2}}$ and $\ket{70S_{1/2}}$. The calculated pair-interaction potential for the state pair $\ket{70P_{1/2}, m_j = 1/2} \otimes \ket{70S_{1/2}, m_j = 1/2}$ exhibits both attractive and repulsive regions, as shown in Fig.~\ref{fig:2}(c). At an interatomic separation of $r = 5.8\ \si{\micro\meter}$, relevant to our experimental conditions, the van der Waals interaction energy compensates a laser detuning of $\Delta = +50$ MHz, thus fulfilling the facilitation condition for interstate excitation.

To probe this regime experimentally, we first prepare a small number of Rydberg atoms, approximately $1.25$ seeds on average, in the $\ket{70P_{1/2}}$ state, and subsequently drive facilitated excitation between ${70S_{1/2}}$ levels. We then repeat the experiment without the initial seeds in the $\ket{70P_{1/2}}$ state.

Figure~\ref{fig:2}(a)-(b) presents the facilitated excitation dynamics as a function of the excitation time. The number of additional $\ket{70S_{1/2}}$ excitations induced by the presence of $\ket{70P_{1/2}}$ seeds reaches $\Delta N \approx 4$ after $100\,\si{\micro\second}$. Simultaneously, the Mandel $Q$ parameter increases significantly, attaining values up to $Q \approx 5$ at $t_{\text{exc}} = 130\, \si{\micro\second}$. These observations confirm the onset of strong correlated excitation dynamics indicating the occurrence of interstate facilitation.

\section{Conclusion}

We have experimentally demonstrated facilitated
intrastate Rydberg excitation for different pairs of Rydberg
states with different interaction potentials. The nature of the interaction - repulsive or
attractive - dictates the detuning required for facilitated
excitation. Positive detuning corresponds to repulsive
interactions, while negative detuning indicates attractive
interactions. We have also observed signatures of interstate facilitation by using a seed atom in a $70P$ Rydberg level to trigger a facilitation avalanche between $70S$ Rydberg levels.
 
Our work extends the concept of Rydberg facilitation beyond the typically used $nS$ levels to $nP$ and $nD$ levels. This will enable extensions of previously demonstrated phenomena, such as nonequilibirum phase transitions in driven-dissipative systems of strongly interacting Rydberg atoms \cite{rydbergchain,periodicfac,paircorr,phasetransition,fac2dlattice,facchip}, to the case of anisotropic interactions ($P$ and $D$ levels), which will likely give rise to richer dynamics. Our results will also be helpful in designing and investigating systems featuring correlated dissipation through the interplay between engineered dissipation \cite{Kazemi2023, Kitson2024, Reiter2012, Shao2023, Schempp2015, Breuer2007, Kramer2018} and blockade/facilitation effects.

\section{ACKNOWLEDGMENTS}

We gratefully acknowledge funding by the Julian Schwinger Foundation grant JSF-18-12-00, the H2020 ITN “MOQS” (grant agreement number 955479), and the PNRR MUR project PE0000023-NQSTI.

\bibliographystyle{apsrev4-2}
\bibliography{Bibliografia} 

\end{document}